\documentclass[sort&compress]{elsarticle}

\usepackage{lineno,hyperref}
\modulolinenumbers[5]

\usepackage{multirow}
\usepackage{float}

\journal{International Communications in Heat and Mass Transfer}

\usepackage{color} 

\begin{document}

\begin{frontmatter}
\title{Lifetime of sessile saliva droplets in the context of SARS-CoV-2}
\author{Saravanan Balusamy$^{\dagger}$}
\author{Sayak Banerjee$^{\dagger}$}
\author{Kirti Chandra Sahu$^{\dagger\dagger}$\fnref{ksahu@iith.ac.in}}
 \address{
        $^{\dagger}$Department of Mechanical and Aerospace Engineering, Indian Institute of Technology Hyderabad, Sangareddy 502 285, Telangana, India
        $^{\dagger\dagger}$Department of Chemical Engineering, Indian Institute of Technology Hyderabad, Sangareddy 502 285, Telangana, India\\
        }

\begin{abstract}
Spreading of respiratory diseases, such as COVID-19, from contaminated surfaces is dependent on the drying time of the deposited droplets containing the virus. The evaporation rate depends on environmental conditions, such as ambient temperature and relative humidity and physical properties (e.g., droplet volume, contact angle and composition). The respiratory droplets contain salt (NaCl), protein (mucin), and surfactant (dipalmitoylphosphatidylcholine) in addition to water, which are expected to influence the evaporation in a big way.  A diffusion-based theoretical model for estimating the drying time is developed which takes into account the dynamic contact angle of saliva droplets laden with salt and insoluble surfactants. The effect of the initial volume, contact angle, salinity, surfactant concentration, ambient temperature and relative humidity on the drying time of droplets is investigated. 
\end{abstract}


\end{frontmatter}

\section{Introduction} \label{sec:intro}

SARS-CoV-2 (also known as coronavirus) has spread to more than 200 countries and infected more than 11 million people worldwide until the first week of July 2020. Respiratory infections such as SARS-CoV-2 virus spreads primarily by respiratory droplets (size $> 10 \mu$m) of saliva or discharge from the nose of an infected person during coughing/sneezing \cite{li2020early,tang2006factors}. The study of fluid dynamics of respiratory droplets migrating in the air or deposited on surfaces helps to understand the life of these droplets \cite{mittal2020flow,bhardwaj2020likelihood}. Few researchers (see, for instance, Ref. \cite{bourouiba2020turbulent}) studied the distance travelled by droplets and their lifetime (drying time) in the air under various respiratory conditions. Chen \cite{chen2020effects} studied the effect of ambient temperature and humidity on the life time of sneeze droplets using one-dimensional droplet evaporation model. The World Health Origination (WHO) has suggested several guidelines, such as wearing face masks and maintaining social distance as the preventive measures for air transmission of COVID-19. 

The respiratory droplets substantially increase their lifetime as they come into contact with a surface depending on the surface properties and local ambient conditions. Bhardwaj and Agrawal \cite{bhardwaj2020likelihood} recently used a diffusion-based model to study pinned droplet evaporation with a constant contact angle assuming that SARS-CoV-2 droplet is pure water. They also remarked that the quasi-steady assumption used in their study is valid when the heat equilibrium time is much smaller than the drying time of the droplet \cite{larson2014transport}. Respiratory droplets, however, also contain salt (NaCl), protein (mucin), and surfactant (dipalmitoylphosphatidylcholine), which are supposed to influence the drying time. In particular, the concentration of NaCl, mucin and surfactant per litre of saliva solution are 9 g/l, 3 g/l and 0.5 g/l, respectively \cite{vejerano2018physico}. Thus, in this study, we have extended the study of Bhardwaj and Agrawal \cite{bhardwaj2020likelihood} to investigate the lifetime of sessile droplets laden with salinity and other insoluble ingredients (surfactant) on various surfaces under different atmospheric conditions, such as relative humidity (RH) and temperature ($T$), which has been poorly understood until now.

In the context of normal droplets, the evaporation has been investigated theoretically and experimentally by many researchers (see e.g. Refs. \cite{bhardwaj2020likelihood,gurrala2019evaporation,fukatani2016effect,sefiane2003experimental}). Gurrala et al. \cite{gurrala2019evaporation} investigated the evaporation of sessile droplets at different temperatures. They compared the results obtained from the theoretical model that includes the diffusion, free convection and passive transport of air with their experiments for different compositions and surface temperatures. Fukatani et al. \cite{fukatani2016effect} experimentally studied the effect of the relative humidity and ambient temperature on the early stages of the evaporation of organic solvents. Karapetsas et al. \cite{karapetsas2016evaporation} investigated the evaporation of sessile droplets laden with particles and insoluble surfactants using lubrication theory. They found that the lifetime of the droplet is significantly influenced by the presence of the surfactant via the spreading of the droplet and Marangoni convection. Souli{\'e} et al. \cite{soulie2015evaporation} compared the initial evaporation rate (linear stage) of droplets of aqueous saline (NaCl) solutions with that obtained from their experiments, and found that the evaporation rates for high salt concentrations and small contact angles were significantly lower than those predicted from the well-accepted diffusion-controlled evaporation scenario for sessile droplets. They also observed that the droplet remains pinned when the concentration of the salt is $\ge 10^{-6}$ mole. 

\section{Mathematical modelling} \label{sec:model}

For small droplets (volume, $V \le \mathcal{O}(1)$ $\mu$L) , the surface tension force predominates over the gravitational force and a sessile droplet obtains a symmetrical spherical cap shape \cite{brutin2018recent}. For such a droplet, if $A (t) = \theta(t) \cdot \pi/180$ is the contact angle in radians and $R$ is the wetting radius of the droplet, then the time-varying droplet volume can be expressed as
\begin{equation} 
V (t) = {\pi R^3 \over 3} {(1-\cos A(t))^2 (2 + \cos A(t)) \over \sin^3 A(t)}.\label{eqn 1}
\end{equation}

We make the following assumptions for the deduction of the rate of evaporation of droplets laden with salinity and insoluble surfactants. (i) The surface is at the same temperature as the surrounding air, (ii) the temperature remains constant during the evaporation process, (iii) the droplet remains internally well-mixed during the evaporation process and (iv) the Stefan flow and the effect of solutal Marangoni convection inside the droplet are neglected. These assumptions are valid for the evaporation of small droplets. Under these conditions, one can ignore the convective mass transfer and assume that the droplet evaporates at the vapour-liquid interface only through the processes driven by diffusion. Therefore, the well studied expression \cite{hu2002} of the evaporation rate, ${\dot m}$ (kg/s) of spherical cap-shaped droplets through the diffusion process are adopted in this work, which is given by
\begin{equation}
{\dot m} =  - \pi R {\cal D} (T) C_{sat} (T)\left[1 - {\rm RH}\right] f( A) \alpha, \label{eqn3}
\end{equation}
where {\rm RH} is the relative humidity of air (\%) and $f(A) = 1.3 + 0.27 A^2$ for $\theta \le 90^\circ$ \cite{hu2002}. The diffusion coefficient, ${\cal D} (T)$ of water vapour ($m^2/s$) at the given temperature, $T$ ($^\circ$C) is given by \cite{bhardwaj2020likelihood}
\begin{equation}
{\cal D}(T) = 2.5\times10^{-4} \exp \left (-684.15 \over T+ 273.15 \right).\label{eqn4}
\end{equation} 

For water-salt-surfactant solutions, the saturation vapour density, $C_{sat}$, of  water (kg/m$^3$) over the solution interface can be obtained from the Raoult's law,
\begin{equation}
C_{sat} = X_w C^\circ_{sat}, \label{eqn 5}
\end{equation}
where $X_w$ is the mole fraction of water in the solution, and $C^\circ_{sat}$ is the saturation vapour density of pure water at the given temperature, which is given by \cite{bhardwaj2020likelihood} 
\begin{eqnarray}
C^\circ_{sat} (T) = 9.99 \times10^{-7} T^3 - 6.94 \times10^{-5} T^2 +  \nonumber \\ 3.20 \times10^{-3} T - 2.87 \times10^{-2}.
\end{eqnarray} 

In Eq. (\ref{eqn3}), $\alpha$ is the accommodation coefficient for evaporation of the droplet laden with insoluble surfactants \cite{karapetsas2016evaporation}, which is given by
\begin{equation}
\alpha= {1 \over {1 + \Psi {\Gamma \over \Gamma_\infty}}},
\end{equation}
where, the surfactant parameter, $\Psi \ge 0$, and 
\begin{eqnarray}
\Gamma= {\hat{n}_{sur} \over \hat{n}_w(t)+ \hat{n}_{s}+\hat{n}_{sur}}, \quad 
\Gamma_\infty = {\hat{n}_{sur} \over  \hat{n}_{s}+\hat{n}_{sur}}.
\end{eqnarray}
Here, $\Gamma$ is the concentration of the surfactant and $\Gamma_\infty$ is the maximum possible surfactant concentration which is assumed to be the concentration corresponding to a fully dried drop. Also, $\hat{n}_{sur}$ is the number of moles of surfactant, $\hat{n}_{s}$ denotes the number of moles of solute ion particles and $\hat{n}_w(t)$ is the number of moles of water in the droplet at the given time. The number of moles of surfactant is calculated using initial value of $\Gamma=1.22 \times 10^{-5}$ \cite{vejerano2018physico}. The mole fraction of the water in the solution decreases as the evaporation process progresses over time. This decrease in the mole fraction of water in the evaporating droplet increases the surfactant and salt concentrations which, along with a decrease in the contact angle, slows down the rate of evaporation as time progresses. At any given time, the mole fraction of liquid water that remains in the solution can be calculated as
\begin{equation}
X_w(t) = {\hat{n}_w(t) \over \hat{n}_w(t)+ \hat{n}_{s}+\hat{n}_{sur}}. \label{eqn6}
\end{equation}

Note that for electrolyte solutes that dissociate into ions, the sum of the number of moles of the positive and negative ions formed by the solute is equal to $\hat{n}_{solute}$. The molality $(M)$ is defined the ratio between the number of moles of solute ($\hat{n}_{solute}$) and mass of water ($m_w$ in kg)  in the droplet solution. For a salt, like NaCl, $\hat{n}_{s} = 2\hat{n}_{solute}$. Since salt and surfactants are non-volatile, the mass evaporation rate is entirely due to the loss of water. We neglect any loss of these ingredients due to crystallisation from the solvent as the initial droplet is very dilute and no deposition is expected till the very end of the droplet's lifetime. Thus the rate of change of the number of moles of water from the droplet solution can be expressed as
\begin{equation}
\frac{d\hat{n}_w}{dt} = \frac{{\dot m}}{{\cal M}_{w}}, \label{eqn7}
\end{equation}
where ${\cal M}_w$ is the molecular weight of water. The rate of change in the droplet volume is given by
\begin{equation}
\frac{dV}{dt} = \frac{{\dot m}}{\rho(t)}, \label{eqn8}
\end{equation}
where $\rho (t)$ is the density of the solution at a given time. The fluid density is a function of the temperature and the salt and surfactant concentrations in the solution and its value is obtained from polynomial fits of appropriate thermodynamic data charts \cite{PerryHandbook}. 

\section{Results and discussion}
\label{sec:dis}

The conditions used in the present study are summarised below. The initial contact angle ($\theta_0$) of the droplet is varied from $10^\circ-90^\circ$ to represent different surfaces used in our everyday activities. Note that the contact angle of a droplet on glass, stainless steel, cotton, wood, smartphone screen and coated paper are about $10^\circ$ \cite{roux2004dynamics}, $30^\circ$ \cite{chandra1991collision}, $50^\circ$ \cite{hsieh1996water}, $65^\circ$ \cite{mantanis1997wetting}, $85^\circ$ \cite{jang2014preparation}, and $90-110^\circ$\cite{pykonen2010evaluation}, respectively. The relative humidity (RH) and temperature ($T$) are varied from 10\% - 90\% and $20^\circ$C - $50^\circ$C, respectively. Such ambient conditions are observed in various places worldwide. The initial volume of the droplet $(V_0)$ is varied from 1 nL - 10 nL. In addition, the molality $(M)$ of the saliva is fixed at $0.154$ mol/kg \cite{vejerano2018physico}. For a typical value of $\Gamma/\Gamma_\infty$, the value of $\Psi$ varies from 0 - 50 for different realistic values of $\alpha$ \cite{miles2012comparison}.

In the present study, given the initial mole numbers of water, surfactant and solute ions in the droplet, an iterative numerical code is developed which calculates the mass loss rate while dynamically correcting the changes induced by the decrease in the contact angle and the increase in the salt and surfactant concentrations in the liquid during the evaporation process. 


\begin{figure}
\centering
\hspace{0.1cm} ({\large a}) \hspace{5.0cm} ({\large b}) \\
\includegraphics[width=0.45\textwidth]{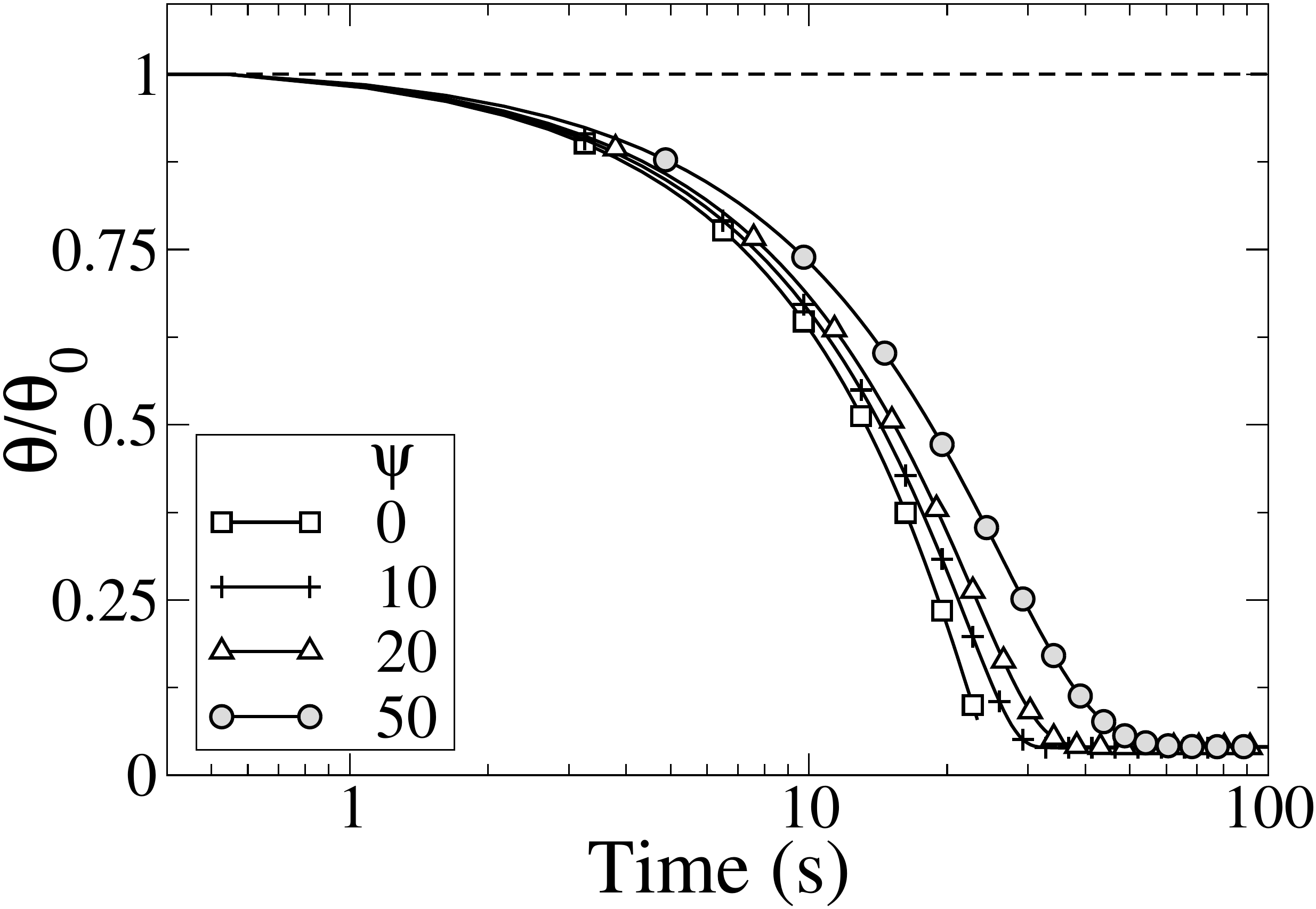} \hspace{0mm} \includegraphics[width=0.45\textwidth]{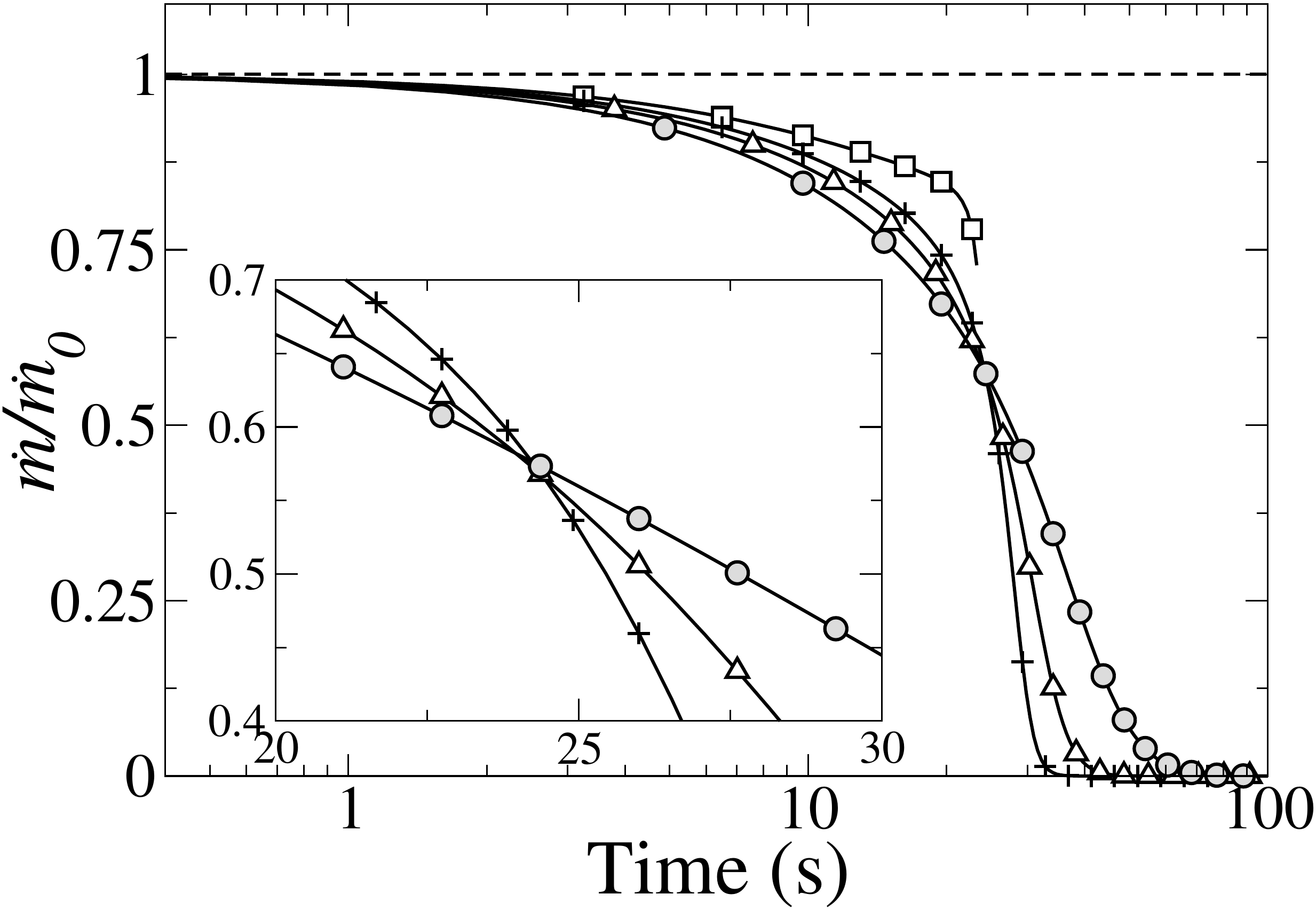}
\caption{Temporal variations of (a) the contact angle of the droplet normalised with the initial contact angle, $\theta/\theta_0$, and (b) the normalised instantaneous mass evaporation rate of the droplet with its initial mass evaporation rate, ${\dot m}/{\dot m_0}$ for different values of $\Psi$. The rest of the parameters are $V_0=10$ nL, $M=0.154$ mol/kg, $T=30^\circ$C, RH = 50\% and $\theta_0 = 50^\circ$. The inset in panel (b) shows a magnified view during the crossover. The dashed horizontal line represents the constant contact angle case \cite{bhardwaj2020likelihood}.}
\label{drying_time2}
\end{figure}

We begin the presentation of our results by examining the effect of the surfactants on evaporation by exploring the influence of the surfactant parameter $\Psi$. Figs. \ref{drying_time2}(a) and (b) show the temporal evolutions of the normalised contact angle (${\theta}/{\theta_0}$) and the normalised instantaneous mass evaporation rate $({\dot m}/{\dot m_0})$ of the droplet for different values of $\Psi$, respectively. The rest of the parameters are $V_0=10$ nL, $M=0.154$ mol/kg, $T=30^\circ$C, RH = 50\% and $\theta_0 = 50^\circ$. Souli{\'e} et al. \cite{soulie2015evaporation} found that a saline droplet remains pinned when the concentration of the salt is $\ge 10^{-6}$ mole. Thus, we simulate the evaporation of a pinned droplet whose contact angle decreases during the evaporation. The black dashed line in Figs. \ref{drying_time2}(a) and (b) shows the constant contact angle case, which does not represent the exact nature of a pinned droplet. The profiles of the normalized contact angle and the normalized mass loss rate show three distinct regions: an initial period of gradual decrease followed by a steep decline at the intermediate period and an eventual near-horizontal line at very small values till the end of the droplet lifetime. It can be seen in Fig. \ref{drying_time2}(a) that the rate of decrease of the contact angle is slower as we increase the value of $\Psi$. For $\Psi=0$, which corresponds to the case when there is no surfactant effect on the evaporation rate, the contact angle of the droplet approaches zero more quickly than for $\Psi>0$, i.e. the droplet becomes a thin film. Inspection of Fig. \ref{drying_time2}(b) reveals that increasing $\Psi$ decreases the rate of evaporation of the droplet at an early stage, while it increases significantly at the intermediate stage as the rate of decline becomes less steep with increasing $\Psi$. The crossover takes place at $t \approx 24$ for the set of parameters considered in the present study (see, inset in Fig. \ref{drying_time2}b). There is an overall increase in the lifetime of the droplet with an increasing value of $\Psi$. This is due to the competition between the decrease in dynamic contact angle and increase in $\Psi$. Unlike, what is expected in the constant contact angle case \cite{bhardwaj2020likelihood}, the normalised rate of mass evaporation decreases with time due to the shrinking free surface (liquid-air interface; $f(A)$ in Eq. \ref{eqn3}) area of the droplet at the early stage. The effect of surfactant parameter is more pronounced at the late stage of evaporation, as the mole fractions of the salt and surfactant significantly increase in the droplet (Fig. \ref{drying_time2}(b)). 
Both the Figs. \ref{drying_time2} (a) and (b) show a highly nonlinear evolution of the contact angle and the instantaneous mass-loss rates that are significantly affected by the surfactant spreading coefficient values.

\begin{figure}
\centering
\hspace{0.1cm} ({\large a}) \hspace{5.0cm} ({\large b}) \\
\includegraphics[width=0.45\textwidth]{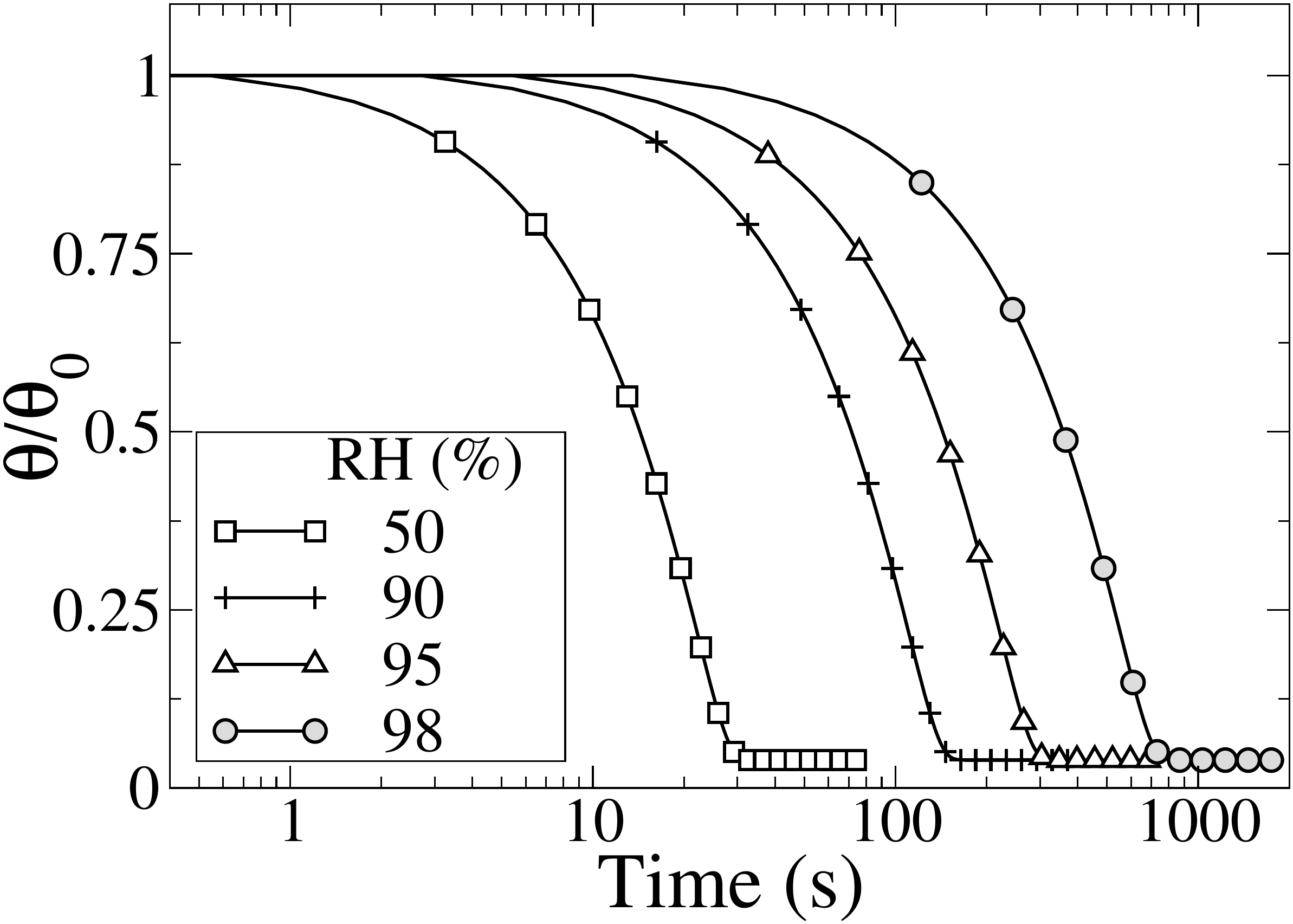} \hspace{0mm} \includegraphics[width=0.45\textwidth]{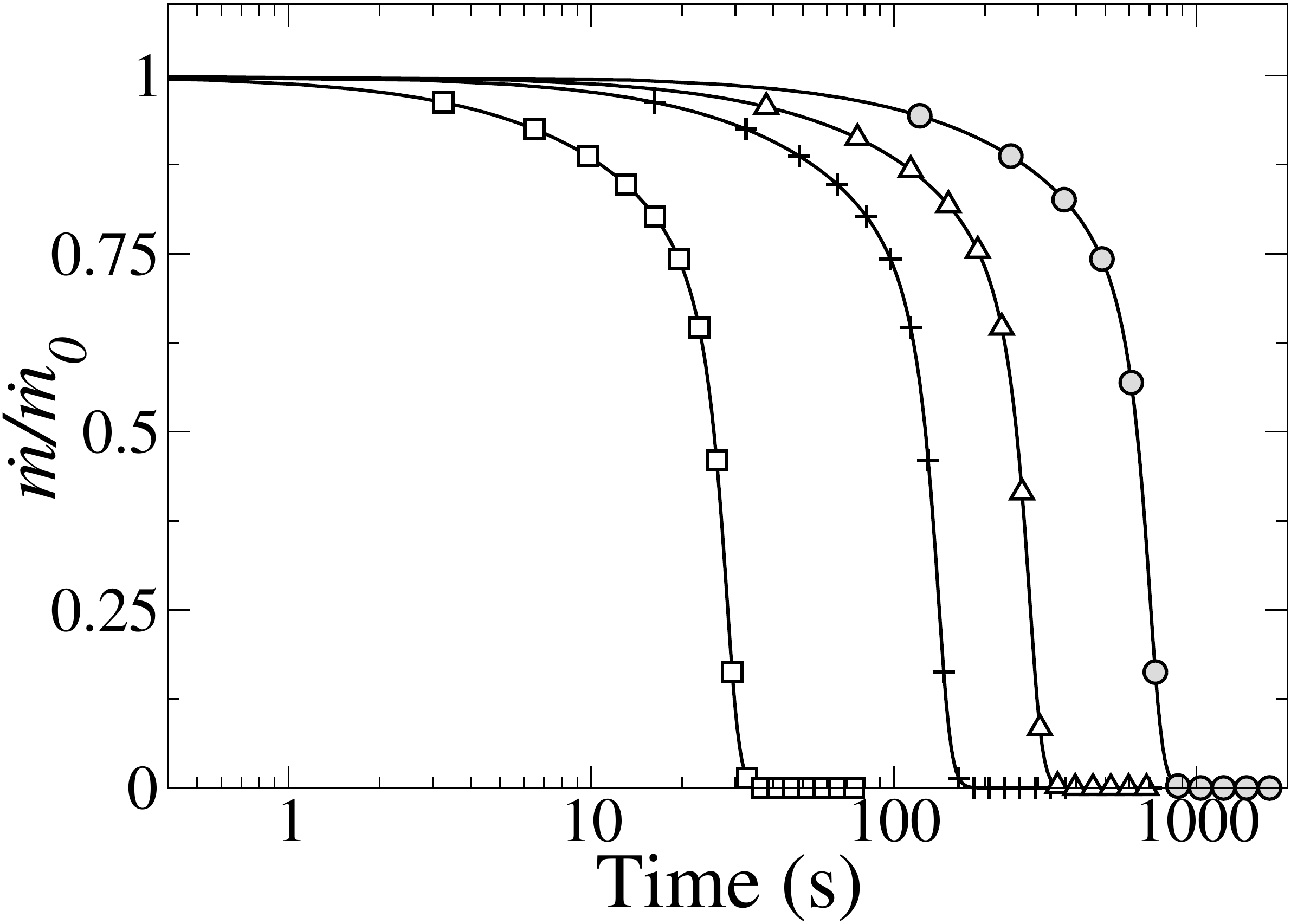} 
\caption{Temporal variations of (a) the contact angle of the droplet normalised with the initial contact angle, $\theta/\theta_0$, (b) the normalised instantaneous mass evaporation rate of the droplet with its initial mass evaporation rate, ${\dot m}/{\dot m_0}$ for different values of RH (\%). The rest of the parameters are $V_0=10$ nL, $M=0.154$ mol/kg, $T=30^\circ$C, $\Psi = 10$ and $\theta_0 = 50^\circ$.} 
\label{drying_time1}
\end{figure}

\begin{figure}
\centering
\hspace{0.0cm} ({\large a}) \hspace{5.0cm} ({\large b}) \\
\includegraphics[width=0.45\textwidth]{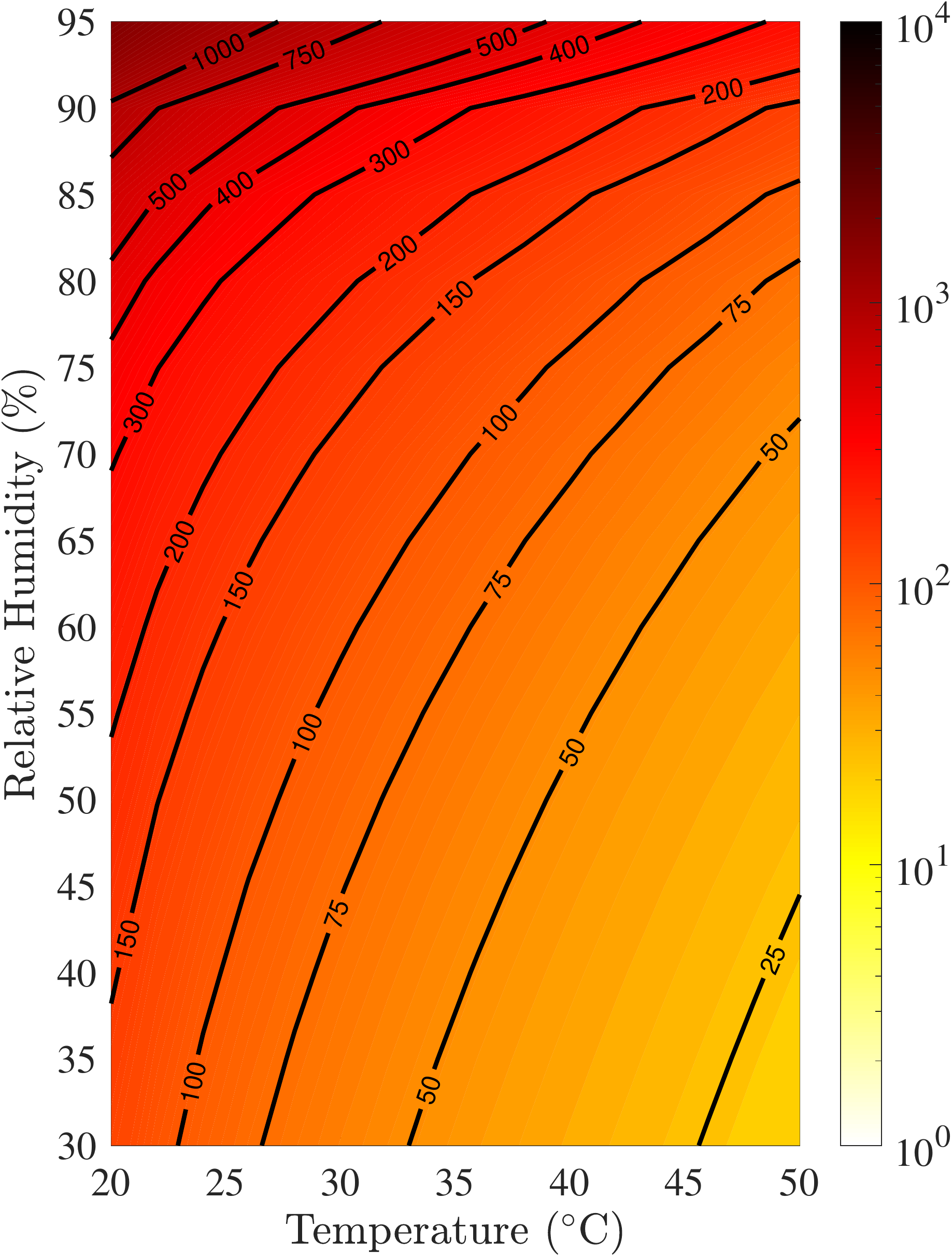} \hspace{0mm} \includegraphics[width=0.45\textwidth]{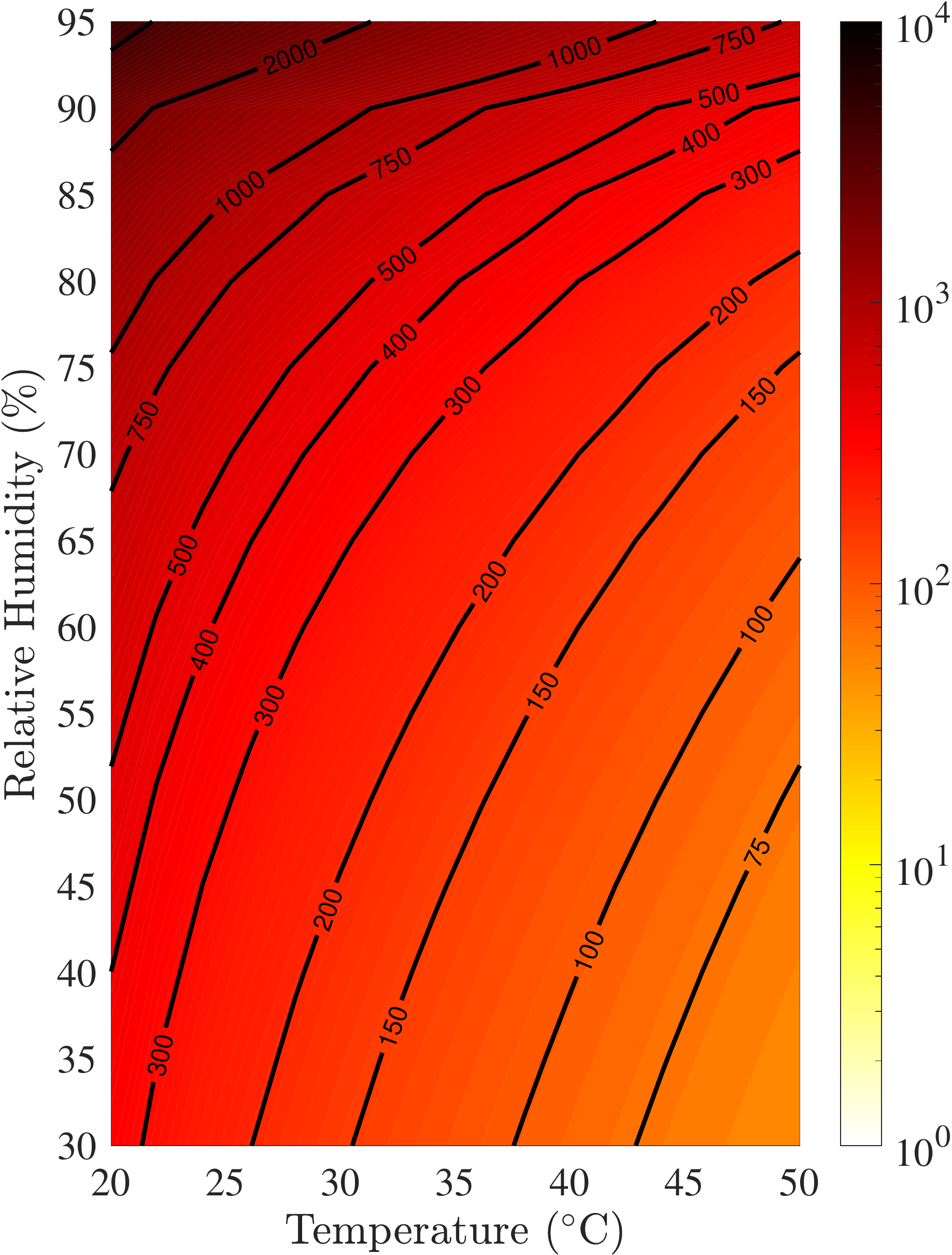} \\
\caption{Regime maps showing the droplet's drying time in the $T-$RH space. (a) $\theta_0=10^\circ$,  (b) $\theta_0=90^\circ$. The colorbar shows the drying time in second in the logarithmic scale. The rest of the parameters are $V_0=10$ nL, $M = 0.154$ mol/kg and $\Psi = 20$.}
\label{regime_map}
\end{figure}

Another important factor influencing the drying time of the droplets is the relative humidity. In Fig. \ref{drying_time1}(a), it can be seen that increasing RH (\%) decreases the rate of decline of the normalised contact angle $(\theta/\theta_0)$. This is because increasing RH decreases the evaporation rate of the pinned droplet as shown in Fig. \ref{drying_time1}(b). It is found (not shown) that the mole fraction of the water remains close to one for a longer time as the value of RH is increased and subsequently, decreases rapidly. These distinct stages are apparent in Fig. \ref{drying_time1}(b), which shows that in the early stage, the normalised instantaneous mass evaporation rate, ${\dot m}/{\dot m_0}$ is almost constant, and is followed by a rapid decrease in ${\dot m}/{\dot m_0}$ in the late stage. It can also be seen in Fig. \ref{drying_time1}(b) that the lifetime of the droplet increases significantly for RH $\ge 90$\%. This range of humidity corresponds to the monsoon scenario in the Indian subcontinent where the RH peaks at around 90\% - 100\%. We found that the droplet lifetimes are approximately equal to 32.5 seconds for RH $= 50$\%, 180 seconds for RH $= 90$\%, 350 seconds for RH $= 95$\% and 860 seconds for RH $= 95$\%. These values demonstrate the high sensitivity of the droplet lifetime on the relative humidity of the ambient. 

To further elucidate the effect of the relative humidity, temperature, initial contact angle and $\Psi$ on the lifetime of droplets, we present the regime maps in Figs. \ref{regime_map}(a) and (b), showing the drying time of the droplet in $T -$ RH space in the two limits of initial contact angle considered in the present study, namely, $\theta_0=10^\circ$ and $\theta_0=90^\circ$, respectively for $V_0=10$ nL, $M = 0.154$ mol/kg and $\Psi = 20$. 
The isocontour lines in these maps show the drying time in second. A logarithmic scale is used for the colourmap to represent the drying time of the droplet as it varies by 4 orders of magnitude. The longest drying times are observed with the combination of low $T$ and high RH (upper left corner), while the drying time progressively reduces as the humidity falls and the temperature rises for both sets of parameters considered. We also found that for a fixed initial volume of the droplet, increasing the initial contact angle increases the drying time. This follows physical intuition, as a droplet with a smaller contact angle has a greater wetting radius and surface area for the same volume resulting in a larger interface area over which the diffusion-driven evaporation can occur. Thus, highly hydrophilic surfaces may be less susceptible to prolonged contamination in comparison to less hydrophilic surfaces, which is a somewhat counter-intuitive result. Increasing the surfactant parameter, $\Psi$ also found to increase the drying time of the droplet.

\begin{figure}
\centering
\hspace{0.1cm} ({\large a}) \hspace{5.0cm} ({\large b}) \\
\includegraphics[width=0.45\textwidth]{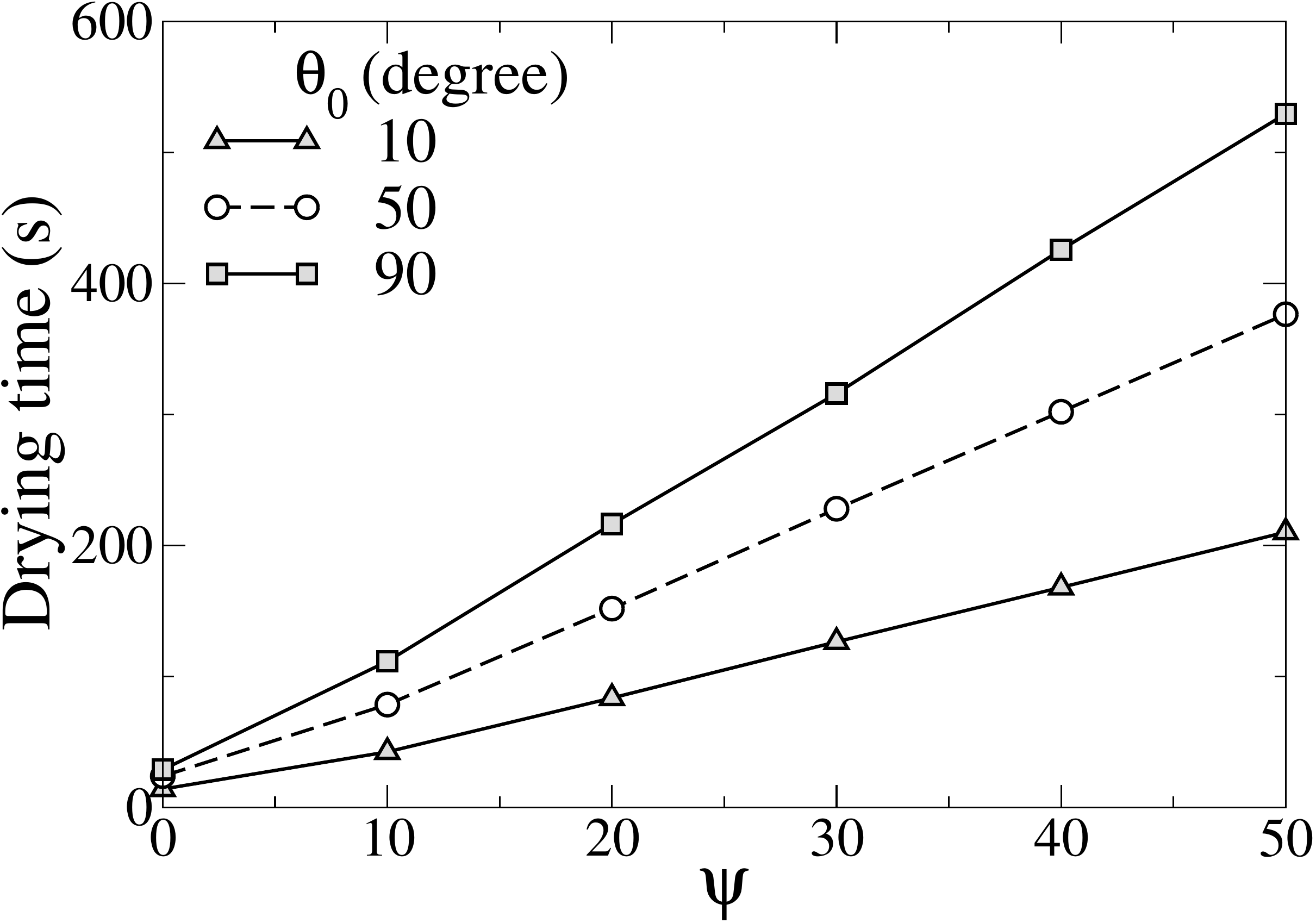} \hspace{0mm} \includegraphics[width=0.45\textwidth]{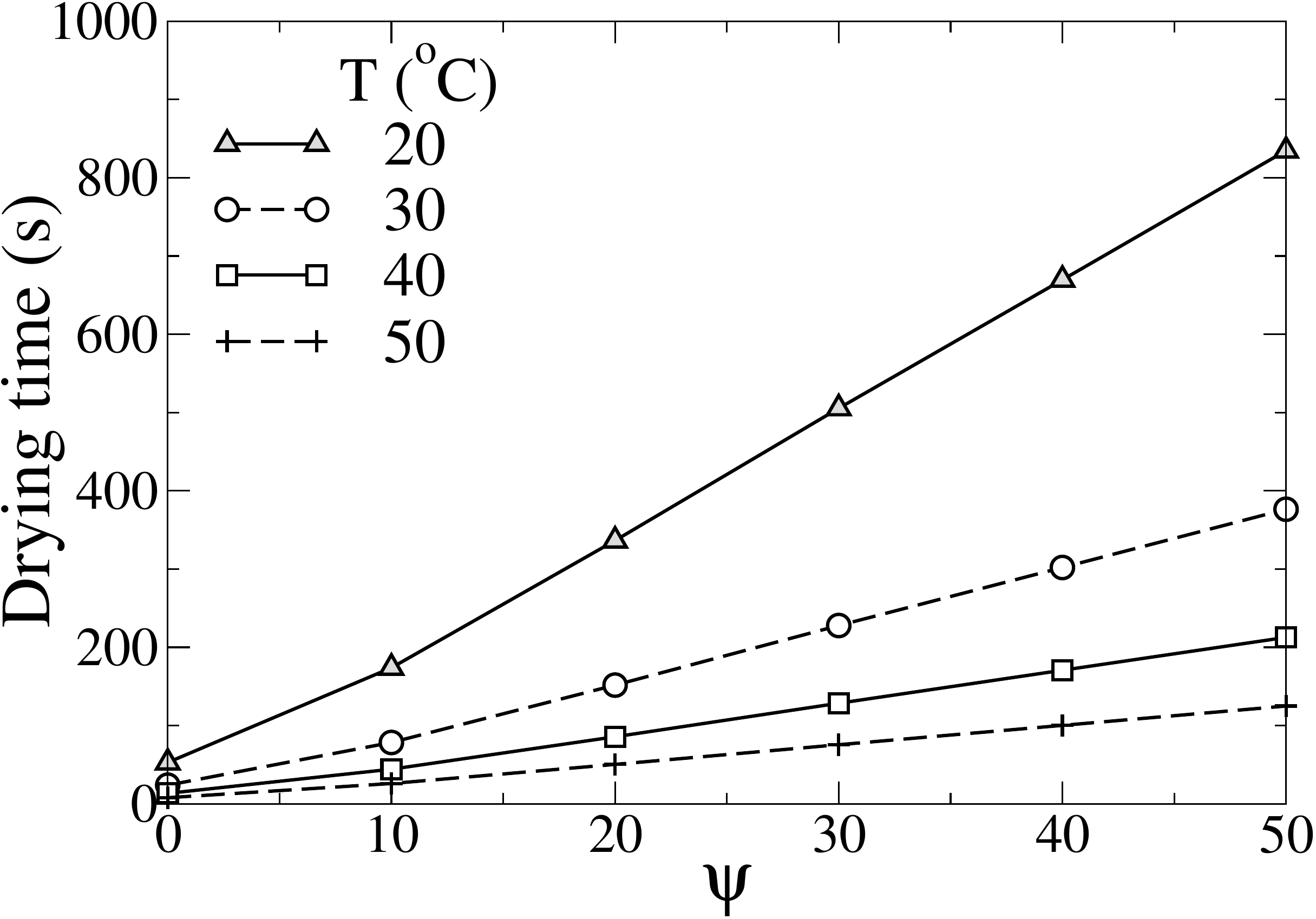} \\
\hspace{0.1cm} ({\large c}) \hspace{5.0cm} ({\large d}) \\
\includegraphics[width=0.45\textwidth]{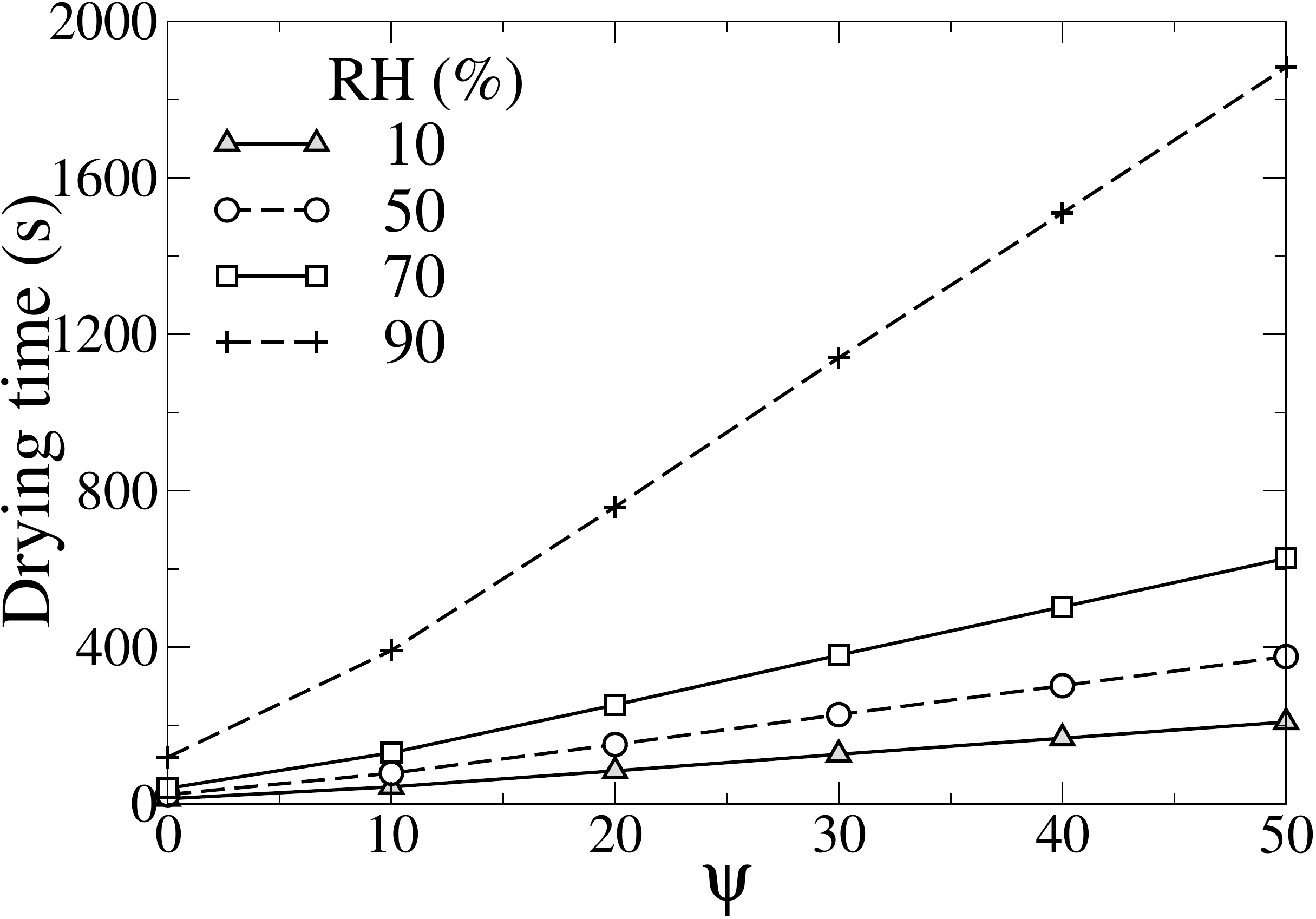} \hspace{0mm} \includegraphics[width=0.45\textwidth]{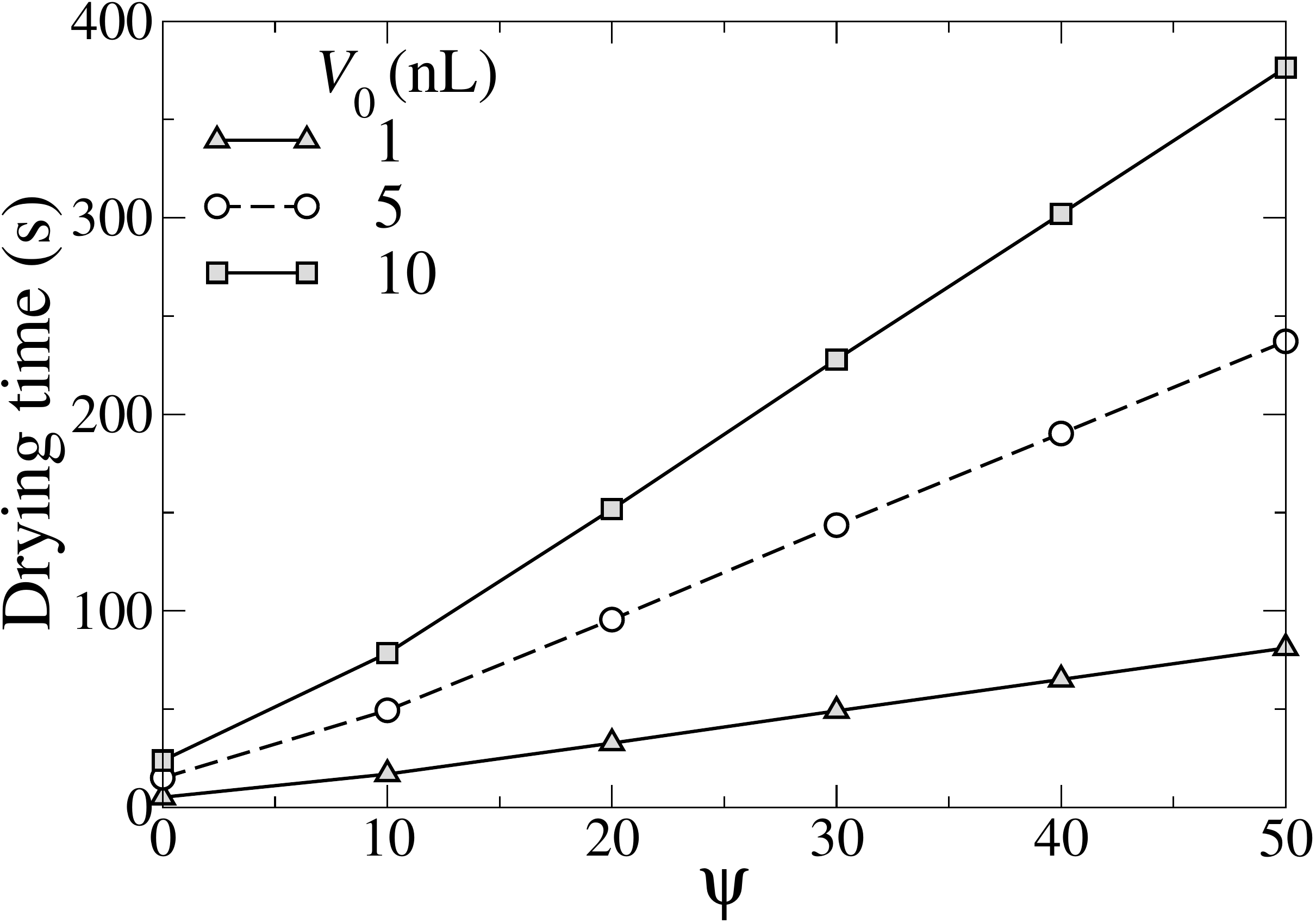} \\
\caption{Variation of the drying time of the droplet with $\Psi$ for $M=0.154$ mol/kg. (a) Effect of $\theta_0$ (degree) for $V_0=10$ nL, $T=30^\circ$C and RH = 50 \%. (b) Effect of $T$ ($^\circ$C) for $V_0=10$ nL, RH = 50 \% and $\theta_0 = 50^\circ$. (c) Effect of RH (\%) for $V_0=10$ nL, $T=30^\circ$C and $\theta_0 = 50^\circ$. (d) Effect of $V_0$ (nL) for $T=30^\circ$C, RH = 50 \% and $\theta_0 = 50^\circ$.}
\label{drying_time}
\end{figure}

Finally, in Figs. \ref{drying_time}(a-d), we present the variations of the lifetime of the saliva droplets laden with salt and insoluble surfactants with $\Psi$ for different values of $\theta_0$, $T$, RH and $V_0$, respectively. It can be seen that increasing $\Psi$, $\theta_0$ and RH and $V_0$ and decreasing $T$ increases the lifetime of the droplet for a fixed set of other parameters. It can be seen that the drying time increases with increasing $\Psi$ for all cases. Further, the rate of increase of the drying time with $\Psi$ becomes larger as the value of the initial contact angle, relative humidity and initial droplet volume increases and the value of the ambient temperature falls. Thus the surfactant induced decrease in the rate of evaporation has a more pronounced effect in enhancing the lifetime of the droplet especially when the basal evaporation rate is slow due to the other conditions, such as high RH, low $T$, higher $\theta_o$ and higher $V_0$.

\section{Concluding remarks}
\label{sec:conc}

To summarise, in our study, a diffusion-based theoretical model is developed which takes into account the dynamic contact angle, the salt concentration and the surfactant effect during the evaporation of sessile saliva droplets and the associated physics. The current work is an extension of Bhardwaj and Agrawal's study \cite{bhardwaj2020likelihood} to consider saliva properties (such as salinity and other insoluble surfactants) and the dynamic contact angle associated with the evaporating droplet. The effect of the surfactant was modelled by incorporating the accommodation coefficient, $\alpha$ into the diffusion-driven mass evaporation rate. The droplet is assumed to be pinned throughout its lifetime, as is expected for droplets that are saline in nature. We showed that the drying time calculated assuming saliva fluid as water and using a static contact angle significantly under-predicts the real drying time of the saliva droplets. It is observed that drying time is strongly dependent on humidity when RH $> 90$\%. The regime maps in $T-$RH space show that the longest drying time is associated with the combination of low temperature and high RH values. We also found that increasing the initial contact angle and the surfactant parameter, $\Psi$ increases the drying time. The quantitative data of the lifetime of saliva droplets presented in the present study shows the importance of properly sanitising the contact prone surfaces, more so in the monsoon season and air-conditioned indoor areas. \\

\noindent {\bf Data Availability Statement:} The data that support the findings of this study are available from the corresponding author upon reasonable request.

\noindent{\bf Acknowledgement:} {K. C. S. thanks Science \& Engineering Research Board, India for their financial support (Grant number: MTR/2017/000029).}


\end{document}